\documentclass[12pt,aps,pra,superscriptaddress,preprint,a4paper,amsmath,amssymb]{revtex4}

\newcommand{\se}{Schr{\"{o}}dinger equation}

\newcommand{\su}{Surrogate Hamiltonian}
\newcommand{\Op}[1]{{\boldsymbol{\mathrm{\hat{#1}}}}}
\newcommand{\Fkt}[1]{{\mathrm {#1}}}
\newcommand{\sub}[1]{_{\mathrm {#1}}}
\newcommand{\up}[1]{^{\mathrm {#1}}}

\usepackage{graphicx}% Include figure files
\usepackage{dcolumn}% Align table columns on decimal point
\usepackage{bm}% bold math

\begin{document}
\preprint{}

%\begin{frontmatter}


\title{Dissipative quantum dynamics with the {\su} approach.
A comparison between spin and harmonic baths.}

\author{David Gelman}
\email{davg@fh.huji.ac.il}
\affiliation{Fritz Haber Research Center for Molecular Dynamics,\\
Hebrew University of Jerusalem, Jerusalem, 91904, Israel}
\author{Christiane P. Koch}
\email{christiane.koch@lac.u-psud.fr}
\affiliation{Centre National de la Recherche Scientifique, Laboratoire Aim\'{e} Cotton,\\
Campus d'Orsay B\^{a}t. 505, Orsay Cedex, 91405, France}
\author{Ronnie Kosloff}
\email{ronnie@fh.huji.ac.il}
\affiliation{Fritz Haber Research Center for Molecular Dynamics,\\
Hebrew University of Jerusalem, Jerusalem, 91904, Israel}

%%%%%%%%%%%%%%%%%%%%%%%%%%%%%72%%%%%%%%%%%%%%%%%%%%%%%%%%%%%%%%%%%%%%%%%%
\begin{abstract}
The dissipative quantum dynamics of an anharmonic oscillator coupled to 
a bath  is studied with the purpose of elucidating the  differences 
between the relaxation to a spin bath and to a harmonic bath. 
Converged results are obtained for the spin bath by the 
{\su} approach. This method is based on
constructing a system-bath Hamiltonian, with a finite but large number 
of spin bath modes, that mimics exactly a bath with an infinite number 
of modes for a finite time interval. Convergence with respect to the 
number of simultaneous excitations of bath modes can be checked. 
The results are compared to calculations that include a finite number 
of harmonic modes carried out by using the multi-configuration 
time-dependent Hartree method of Nest and Meyer, 
[J. Chem. Phys. {\bf 119}, 24 (2003)]. In the weak coupling regime, 
at zero temperature and for small excitations of the primary system, 
both methods converge to the Markovian limit. When initially the primary 
system is significantly excited, the spin bath can saturate restricting 
the energy acceptance. An interaction term between bath modes that 
spreads the excitation eliminates the saturation. The loss of phase 
between two cat states has been analyzed and the results for the 
spin and harmonic baths are almost identical. For stronger couplings, 
the dynamics induced by the two types of baths deviate. The accumulation 
and degree of entanglement between the bath modes have been characterized. 
Only in the spin bath the dynamics generate entanglement between the 
bath modes.

\end{abstract}

\date{\today}

%\begin{keyword}

%\PACS
%\end{keyword}
%\end{frontmatter}
\maketitle

%%%%%%%%%%%%%%%%%%%%%%%%%%%%%72%%%%%%%%%%%%%%%%%%%%%%%%%%%%%%%%%%%%%%%%%%

% main text
\section{Introduction}
Modeling quantum many-body systems is  a challenging problem. The
main obstacle is the exponential growth in complexity with the
number of degrees of freedom. Significant simplifications are
achieved by partitioning the total system into a primary part and
a bath describing the environment \cite{weiss}. The
idea is to model the primary system explicitly and the bath
implicitly, thus minimizing the complexity of the bath to its
influence on the primary system. A bath composed of a set of
noninteracting harmonic oscillators is the one most widely used.
The idea originates from a normal mode analysis combined with a
weak system-bath coupling assumption \cite{feynman}. If the bath
is only weakly perturbed by the system, it can be considered
linear, and therefore described as a collection of harmonic
oscillators. Such a bath is natural for systems interacting with
the radiation field \cite{louisell}. The harmonic bath model has
also been applied to less favorable scenarios such as energy
relaxation and dephasing of molecules in the liquid phase or on
solids. In these cases a strong coupling or interactions with a
low-temperature environment may cause large system-bath
correlations, and will therefore result in a failure of the
Markovian approximation. To overcome such difficulties in the
dynamics of molecules that are in intimate interaction with an
environment, an alternative approach termed the {\su} \cite{surr}
has been developed. The {\su} method employs a bath composed of
two-level systems that acts as a spin bath
\cite{caldeira93,makri98,tsonchev,prokofev,hanggi}.

The concept of the system-bath separation underlines the quantum
description of many body dynamics. The origins of the spin and
harmonic baths are different. The harmonic bath is closely related
to a normal mode decomposition. Once this is done the spectral
density function is able to completely determine the relaxation
dynamics. From a computational point of view the determination of
the spectral density is a major task. The most popular working
procedure is to extract it from classical mechanics \cite{berne}. 
The drawback is that this procedure assumes harmonic modes and a 
linear system bath coupling term. The spin bath has its origin 
in a tight binding model of condensed phase. This can also become a
simulation procedure if the parameters of the tight binding model
can be estimated from first principles \cite{ckoch03}.

The purpose of the present study is to compare the performance of
the two baths in a simple system composed of a primary anharmonic
oscillator coupled to a multi-mode bath. In the limit of weak
system-bath coupling, it has been shown that the two baths are
equivalent. For finite temperature the equivalence requires a rescaling
of the spectral density function which determines the coupling of the
primary system to the different bath modes \cite{caldeira93,makri98,tsonchev}.
The limiting coupling strength
where the dynamics induced by the two baths differ has not yet
been characterized. For stronger coupling strength, the ergodic
behavior of the two baths should be different. The bath modes of
the linearly driven harmonic bath are uncorrelated. In the spin
bath the coupling to the primary system induces quantum
entanglement between the different modes. It is valuable to know
how this fundamental difference influences the dynamics of the
primary system.

Our comparative study is based on a numerical model of a system
coupled to a bath, with a large but finite number of  modes. For a
finite interval of time determined by the inverse of the energy
level spacing, the finite bath mimics exactly a bath with an
infinite number of modes. For this interval the primary system
cannot resolve the full density of states of the bath. By
renormalizing the system-bath interaction term to the density of
states, the finite bath faithfully represents the infinite bath up to
this time limit.

The  dynamics of the primary oscillator coupled to the harmonic bath
has been recently calculated based on the multi-configuration time dependent
Hartree approximation (MCTDH) \cite{meyer03,burghardt}. The
authors were able to show that for a Morse oscillator coupled to a
bath, converged results could be obtained for a bath consisting of
60 modes to a time scale of 3 ps. The present study utilized the
same system and system-bath coupling parameters, but employed a
spin bath in the context of the {\su}. The comparison allows an
evaluation of the similarities and differences between the two
descriptions. Once the differences are identified, it becomes
possible to modify the {\su} bath to extend the realm of
similarity.

The system-bath construction in both cases is not Markovian and
differs from the Redfield \cite{redfield1,redfield2} or semigroup
treatments \cite{lindblad,gorini,alicki}. In the weak coupling
limit the numerical study of Nest and Meyer \cite{meyer03} was
able to identify a coupling parameter where the Markovian
semigroup limit was reached. One can reason that the {\su} bath
should behave similarly in this range of coupling parameters.

The present paper is organized as follows: Section
\ref{sec:theory} outlines the theory of the two models: the {\su}
approach which employs a bath of two-level systems (TLS) and the
MCTDH method using an harmonic bath. Section \ref{sec:system}
describes the system used for calculations. Section
\ref{sec:results} compares the results for the two different
environments. The standard process investigated in studies of
quantum dissipative dynamics is energy relaxation (cf. Section
\ref{subsec:energy}). An interaction between bath modes is
introduced in Section \ref{subsec:interr}. The difference between
correlated and uncorrelated initial states is the subject of
Section \ref{subsec:correlation}. In addition, the decoherence in
the TLS bath of the {\su} is compared to that in a bath of
harmonic oscillators (cf. Section \ref{subsec:dephasing}). A
characterization of different kinds of entanglement in the {\su}
approach is presented in Section \ref{subsec:entanglement}.
Finally, Section \ref{sec:conclusions} summarizes and concludes.

Appendix \ref{sec:equations} compares the equations of motion
between the two different types of bath and
Appendix \ref{sec:horod_parameter} introduces two different measures
of entanglement of a two-spin system.
\\
It should be noted that atomic units are used throughout the paper $(\hbar=m_e=a_0=1)$.

\label{sec:intro}

%%%%%%%%%%%%%%%%%%%%%%%%%%%%%72%%%%%%%%%%%%%%%%%%%%%%%%%%%%%%%%%%%%%%%%%%

\section{Theory}
\label{sec:theory}

The system under study describes a primary system immersed in a
bath. The state of the combined system-bath is described by the
wave function $\Psi(\Op{R},\Op{\beta}_1,\ldots,\Op{\beta}_{2^N})$
where $\Op{R}$ represents the nuclear configuration of the
dynamical system, and $\{\Op{\beta}_j\}$ are the bath degrees of
freedom. The Hamiltonian of such a combined system is:
\begin{equation}
\Op{H}=\Op{H}\sub{S}\otimes\Op{I}\sub{B}+\Op{I}\sub{S}
\otimes\Op{H}\sub{B}+\Op{H}\sub{SB}~~.
\label{eq:Htot}
\end{equation}
The primary system Hamiltonian takes the form:
\begin{equation}
\Op{H}\sub{S} = \Op{T} +{V}\sub{S}(\Op{R}) \; ,
\end{equation}
where $ \Op{T}=\Op{P}^{2}/2M$ is the kinetic energy and $V\sub{S}$ is
an external potential, which is a function of the system
coordinate(s) $\Op{R}$. $ \Op{H}\sub{B} $ denotes the bath
Hamiltonian consisting of an infinite sum of single mode
Hamiltonians $\Op h_j$:
\begin{equation}
\Op{H}\sub{B} = \sum_j  \Op{h}_j, \label{eq:Hbath} \; .
\end{equation}
For the harmonic bath the single mode Hamiltonians take the form:
\begin{equation}
\Op {h}_j~~=~~\frac{{\Op p}_j^2}{2m_i} + \frac{m_j
\omega_j^2}{2}\Op{q}_j^2 ~~=~~\omega_j {\Op a}_j^{\dagger}{\Op a}_j \; ,
\label{eq:harm_sing}
\end{equation}
where $\Op {p}_j,\Op {q}_j$ are the normal mode momentum and
coordinate respectively, and $\Op a_j = \sqrt{\frac{m_j\omega_j}{2}}\Op {q}_j +
\frac{i}{\sqrt{2 m_j \omega_j}}\Op {p}_j$ is the corresponding
annihilation operator. For the spin bath:
\begin{equation}
\Op {h}_j ~~=~~\omega_j \Op{\sigma}_j^\dagger \Op{\sigma}_j \; ,
\label{eq:spin_sing}
\end{equation}
where $\Op{\sigma}_j^\dagger$, $\Op{\sigma}_j$ are the standard
spin creation and annihilation operators of mode $j$.

The system-bath interaction $\Op H\sub{SB}$ can be decomposed into
a sum of products of system and bath operators without loss of
generality. Specifically a system-bath coupling inducing
vibrational relaxation is considered:
\begin{equation}
\Op{H}\sub{SB} = - f(\Op{R}) \otimes \sum_j {\Op V}_j \; ,
\label{eq:Hsb}
\end{equation}
where $ {\Op V}_j =  \lambda_j {\Op q}_j = \lambda_j
(\Op{a}_j^\dagger+\Op{a}_j)$ for the harmonic bath and $ {\Op V}_j
= \lambda_j (\Op{\sigma}_j^\dagger+\Op{\sigma}_j)$ for the spin
bath. $f(\Op{R})$ is a function of the system coordinate operator.
The influence of the bath on the primary system is
characterized by the spectral density function $J(\omega)$. To include
the density of states, the definition of the spectral density function
is chosen as \cite{makri99,louisell}:
\begin{equation}
J(\omega) = \sum_{j} |\lambda_j|^2 \rho(\omega)
\delta(\omega-\omega_j) \;, \label{eq:spec}
\end{equation}
that is the system-bath coupling is weighted by density of
states. Thus the constants $\lambda_j$ are determined as:
\begin{equation}
\lambda_j = \sqrt{J(\omega_j)/\rho(\omega_j)} \; , \label{eq:v_i}
\end{equation}
where $\rho(\omega_j)=(\omega_{j+1}-\omega_j)^{-1}$ is the density
of the states of the bath.

Observables associated with operators of the primary system
are determined from the reduced system density operator:\\
$\Op{\rho}\sub{S}(R,R')=\Fkt{tr}_B \left\{ \left | \Psi \left \rangle
\right \langle \Psi \right | \right \}$, where $\Fkt{tr}_B\{~\}$
is a partial trace over the bath degrees of freedom. The
system density operator is constructed from the total system-bath
wave function and only this function is propagated.

Since within a finite interval of time, the system cannot resolve
the full density of bath states, it is sufficient to replace
the bath modes by a finite set. The sampling density in energy of
this set is determined by the inverse of the time interval. The
finite bath of $N$ spins is constructed with a system-bath
coupling term, which in the limit $N\rightarrow\infty$ converges to
the given spectral density of the full bath. The {\su}, as well as
the MCTDH method, consist of a finite number of bath modes, and they are
therefore limited to representing the dynamics of the investigated
system for a finite time (shorter than the Poincar\'{e} period at which
recurrences appear \cite{kampen}). These recurrences are caused by the
finite size of the bath so that after some time the energy flow
into the bath is reflected at its boundaries.

The {\su} contains all possible correlations between the primary
system and the environment. The combined system-bath state is
described by a $2^N$ dimensional spinor with $N$ being the number of
bath modes. The spinor is bit ordered, i.e., the $j$th bit set in
the spinor index corresponds to the $j$th TLS mode, which is excited if the
counting of bits starts at $j=0$. The dimension $2^N$ results from
the total number of possibilities to combine two states $N$ times.
Thus the total wave function can be written as
\begin{equation}
|\Psi(\Op R,\{\Op{\beta_j}\}) \rangle = c_0|\phi_0(\Op R) \rangle +
\sum_{j=0}^{\binom{N}{1}}c_j |\phi_j(\Op R) \rangle +
\sum_{j,k=0}^{\binom{N}{2}} c_{jk}| \phi_{jk}(\Op R)\rangle\ +
\ldots \;,
\label{eq:ci}
\end{equation}
where $\vert\phi_j(\Op R)\rangle = (0,\ldots,\phi_j(\Op
 R),\ldots,0)^T$ is a singly-excited spinor, $\vert\phi_{jk}(\Op
R)\rangle=(0,\ldots,\phi_j(\Op R),\ldots,\phi_k(\Op
R),\ldots,0)^T$ is a doubly-excited spinor and so on. The $j$th
component corresponds to the $j$th TLS being excited.
However, considering all $2^N$ possibilities of combining the bath
modes might not be necessary in a weak coupling limit. In this
case, for short time dynamics, it is possible to
restrict the number of simultaneous bath excitations \cite{k186}.
As an extreme example, only single excitations might be
considered. If one restricts the number of simultaneous excitations,
the dimension of the spinor becomes the sum of
binomial coefficients $\sum_{k=0}^{N\sub{exc}} \binom{N\sub{exc}}{k}$
with $N\sub{exc}$ the number of simultaneous excitations. The construction
is similar to the configuration-interaction (CI) approach in electronic
structure theories. The restriction of simultaneously allowed excitations
leads to significant numerical savings and its validity can be
checked by increasing $N\sub{exc}$.

In the MCTDH method \cite{MCTDH} the wavefunction $\Psi$, which
describes the dynamics of a system with $M$ degrees of freedom, is
expanded as a linear combination of time-dependent Hartree
products:
\begin{equation}
|\Psi(Q_1,\ldots,Q_M,t) \rangle = \sum_{j_1=1}^{n_1} \cdot \cdot
\cdot \sum_{j_M=1}^{n_M} A_{j_1,\ldots,j_M}(t)\prod_{\kappa=1}^M
|\varphi_{j_{\kappa}}^{(\kappa)}(Q_{\kappa},t) \rangle \;,
\label{eq:mctdh_func}
\end{equation}
where $|\varphi_{j_{\kappa}}^{(\kappa)}\rangle$ is the
single-particle function (spf) for the $\kappa$ degree of freedom
and the $A_{j_1,\ldots,j_M}$ denote the MCTDH expansion coefficients.
The total number of coefficients $A_{j_1,\ldots,j_M}$ and
basis function combinations scales exponentially with the number of
degrees of freedom $M$. Considering a system coupled to a
multi-mode bath, the use of the {\it multiconfigurational} wave
function ensures the correct treatment of the system-bath
correlations \cite{cederbaum96,wang}. The method also
enables grouping of several modes together, which reduces both the
number of single-particle degrees of freedom and the correlation
effects between different modes. Although the exact treatment
is contained in the limit of an infinite number of configurations,
in the weak coupling limit, the time-dependent basis employed in
the MCTDH method should be relatively small.
Worth et al. \cite{cederbaum96} have pointed out that even for weak
coupling, one spf per bath mode (the Hartree limit) is not sufficient
to fully describe the system-bath interaction.
However, the number of spf's for the bath degrees of freedom can be
increased until convergence is achieved, which makes this approximation
controllable.

%%%%%%%%%%%%%%%%%%%%%%%%%%%%%72%%%%%%%%%%%%%%%%%%%%%%%%%%%%%%%%%%%%%%%%%%

\section{The model}
\label{sec:system}

The primary system is constructed from an anharmonic (Morse)
oscillator of mass $M$:
\begin{equation}
\Op{H}\sub{S} = \frac{\Op{P}^2}{2M} + D\left (e^{-2 \alpha
\Op{R}}-2e^{-\alpha \Op{R}}\right) \; .
\end{equation}
The coupling term is non-linear in the Morse oscillator coordinate
$R$, but reduces to a linear one for a small $R$:
\begin{equation}
f(\Op{R})=\frac{1-e^{-\alpha \Op{R}}}{\alpha} \; .
\label{eq:fcoup}
\end{equation}
The spectral density function was chosen to be the same as in the
harmonic bath case. For an Ohmic bath the damping rate $\gamma$
is frequency-independent and the spectral density in the continuum limit
is given by
\begin{equation}
J(\omega)= M \gamma \omega
\label{eq:spec1}
\end{equation}
for all frequencies $\omega$ up to the cutoff frequency $\omega\sub{c}$.
A finite bath with equally spaced sampling of the energy range was used.

The parameters used are the same as in Ref. \cite{meyer03}: a well
depth $D$ of $0.018$ a.u., $\alpha=2$ a.u., and a mass of $M=10^5$
a.u. The initial state was chosen to be a Gaussian displaced by
$R_0=2\tilde{R}$ from the origin with a width of
$\sigma=\tilde{R}$ ($\tilde{R}\approx0.09129$ a.u. is the
characteristic length scale of the Morse oscillator). For such a
displacement the coupling term (\ref{eq:fcoup}) is almost linear.
The initial system-bath state has a direct product form where the
bath is at zero temperature. Such a state has no initial
correlations between the system and the bath.

There are a few characteristic time scales of the system. The
period of the Morse oscillator is $\tau\sub{osc}=2\pi/\Omega\approx127$
fs, where $\Omega=\alpha\sqrt{2D/M}$ refers to the harmonic
frequency of the potential. The bath has two time scales.
$\tau\sub{bath}$ is associated with the highest frequency
$\omega\sub{c}=2.5\Omega$ and corresponds to a time scale of 52 fs. The
time scale corresponding to the frequency spacing $\Delta\omega$
defines the Poincar\'{e} period ($\tau\sub{rec}$). It should be larger
than any other time scale of interest. With $\omega\sub{c}$ fixed this
time becomes:
\begin{equation}
\tau\sub{rec}=\frac{2\pi}{\Delta\omega}=\frac{2\pi N}{\omega\sub{c}}\; .
\label{eq:rec}
\end{equation}
Thus, with an increasing number of bath modes, the convergence
progresses in time. In our simulations the number of TLS is chosen
to be $N=20\ldots 60$ (for different coupling strengths), which
ensures that $\tau\sub{rec}$ is greater than the overall simulation time.

The calculations were performed in three different interaction
regimes identified by considering the involved time scales: (i)
weak coupling referring to $\gamma^{-1}=1630$ fs $\gg
\tau\sub{osc},\tau\sub{bath}$; (ii) the intermediate situation
characterized by $\gamma^{-1}=163$ fs $\approx \tau\sub{osc}>\tau\sub{bath}$;
(iii) the strong coupling regime defined by $\gamma^{-1}=54$ fs
$\approx \tau\sub{bath}<\tau\sub{osc}$.

In the simulations discussed below, the average position of the
oscillator and the energy relaxation were calculated for all three
coupling strengths. For comparison, the effective subsystem energy
was defined as in \cite{meyer03}:
\begin{equation}
E\sub{S} = \langle \Op{H}\sub{S}^{'} \rangle = \langle
\Op{H}\sub{S} \rangle\ + 0.5 \langle \Op{H}\sub{SB} \rangle \; .
\end{equation}
It includes half of the system-bath interaction term.

The dynamics of the system combined with the bath is generated by
solving the time-dependent {\se}:
\begin{equation}
\Psi(\Op{R},\{\Op{\beta_j}\},t) = e^{-i\Op{H}t}\Psi(\Op{R},\{\Op{\beta_j}\},0) \;.
\end{equation}
Each spinor component $\psi_j(\Op{R})$ is represented on a spatial
grid. The kinetic energy operator is applied in Fourier space
employing FFT \cite{kosloff}, and the Chebychev method
\cite{tal-ezer} is used to compute the evolution operator.
Numerical details of applying the bath operators have already 
been given in Ref. \cite{surr,ckoch}. 

\section {Results and Discussion}
\label{sec:results}

\subsection{Energy relaxation and small amplitude motion.}
\label{subsec:energy}

%%%%%%%%%%%%%%%%%%%%%%%%%%%%%72%%%%%%%%%%%%%%%%%%%%%%%%%%%%%%%%%%%%%%%%%%
\begin{figure}
\includegraphics[scale=0.5]{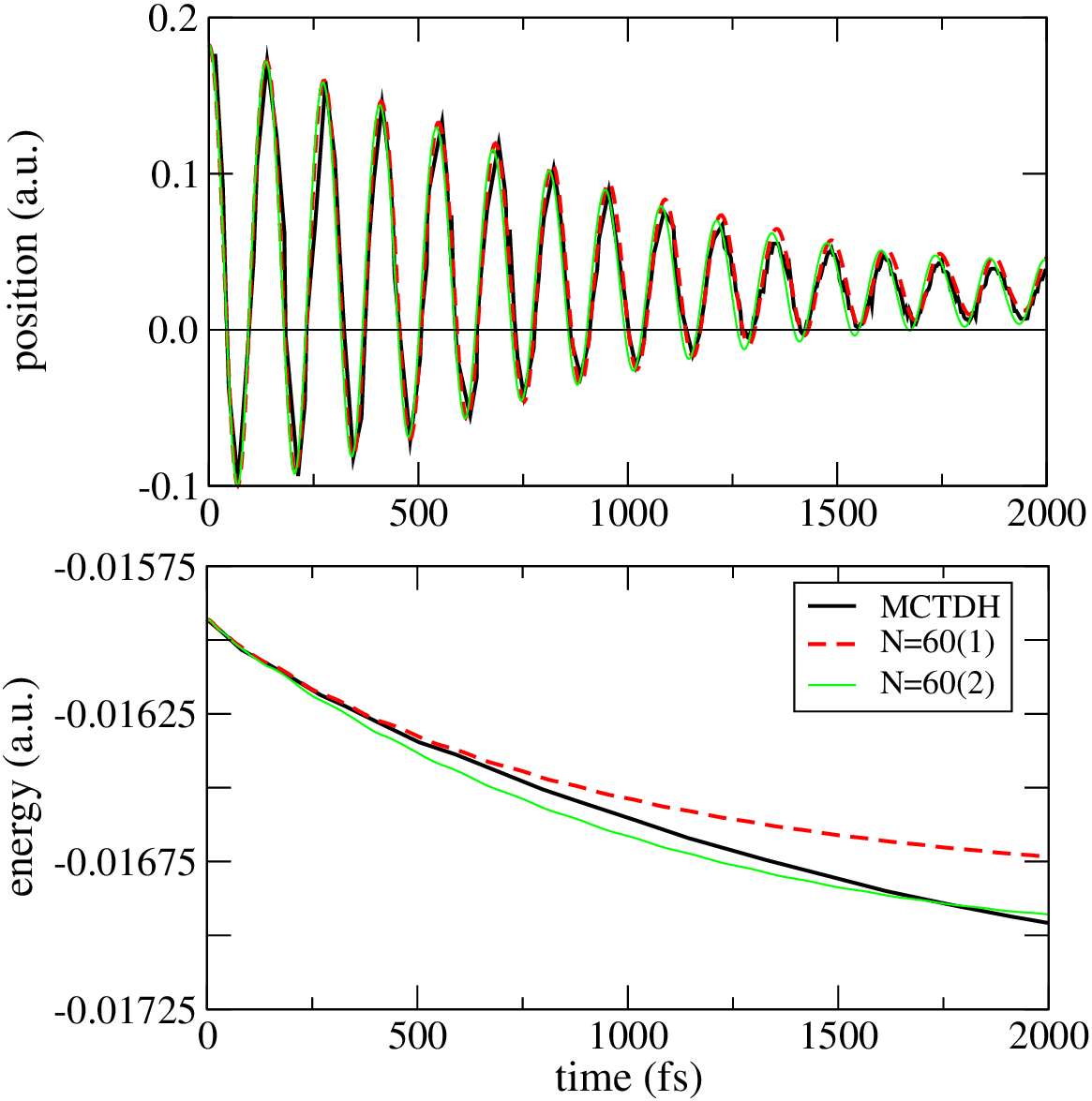}
\caption{\footnotesize The energy relaxation (lower panel) and
damped oscillations of the average position (upper panel) of the
Morse oscillator in the weak coupling limit ($\gamma^{-1}=1630$ fs).
The bath is assumed to be Ohmic with cutoff frequency
$\omega\sub{c}=2.9\cdot10^{-3}$ a.u. and consists of $N=60$ TLS.
The initial state was chosen to be a Gaussian displaced by
$R_0=2\tilde{R}$ with a width of $\sigma=\tilde{R}$, where
$\tilde{R}\approx0.09129$. Thick solid lines refer to MCTDH
calculations with a bath of harmonic oscillators (adopted from
Ref. \cite{meyer03}). Dashed lines refer to {\su} calculations
with only single excitations. Thin lines refer to two simultaneous
excitations allowed.} \vspace{0.8cm} \label{fig:weak}
\end{figure}
%%%%%%%%%%%%%%%%%%%%%%%%%%%%%72%%%%%%%%%%%%%%%%%%%%%%%%%%%%%%%%%%%%%%%%%%
\begin{figure}
\includegraphics[scale=0.5]{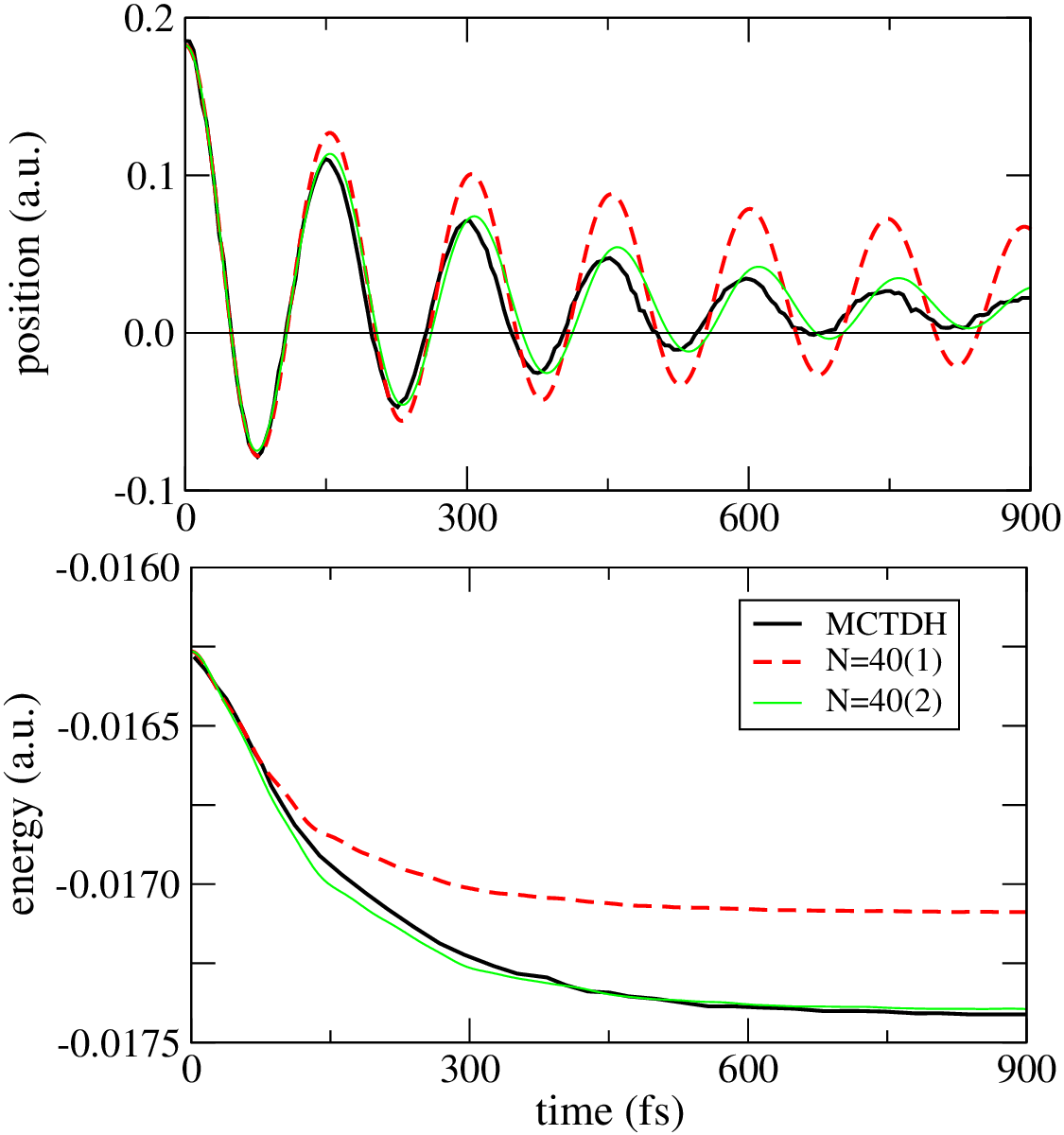}
\caption{\footnotesize The energy relaxation (lower panel) and
damped oscillations of the average position (upper panel) of the
Morse oscillator in the intermediate coupling regime
($\gamma^{-1}=163$ fs). The bath parameters and the initial state
are the same as in the weak coupling calculations. The number of
bath modes is $N=40$. Thick solid lines refer to MCTDH
calculations with a bath of harmonic oscillators (adopted from
Ref. \cite{meyer03}). Dashed lines refer to {\su} calculations
with single excitations only. Thin lines refer to calculations
with two  allowed simultaneous excitations.} \label{fig:medium}
\end{figure}
%%%%%%%%%%%%%%%%%%%%%%%%%%%%%72%%%%%%%%%%%%%%%%%%%%%%%%%%%%%%%%%%%%%%%%%%

%%%%%%%%%%%%%%%%%%%%%%%%%%%%%%%%%%%%%%%%%%%%%%%%%%%%%%%%%%%%%%%%%%%%%%%%%
\begin{figure}
\includegraphics[scale=0.5]{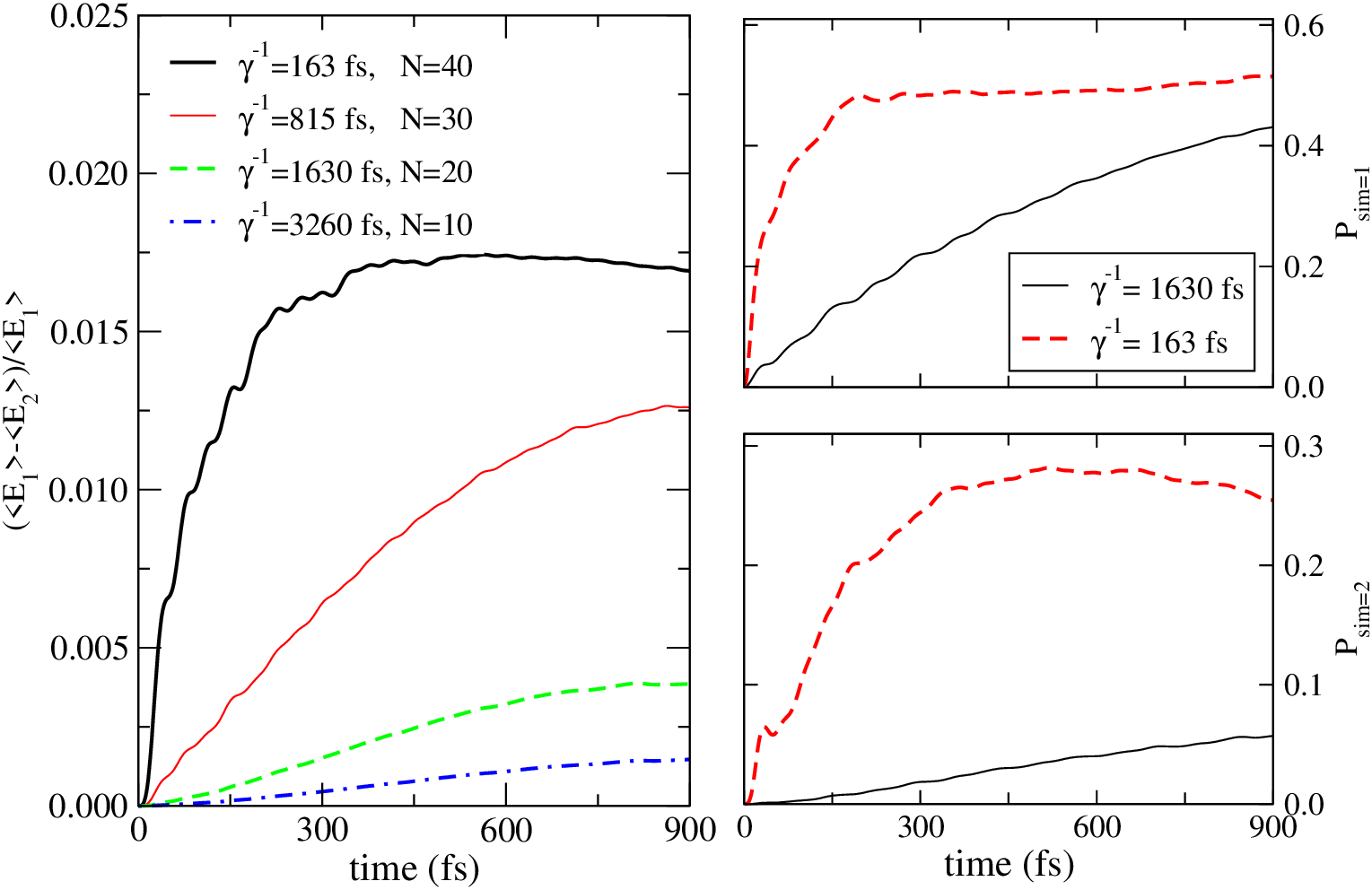}
\caption{\footnotesize (Left panel) The relative difference in
the effective subsystem energy ($\langle H\sub{S} \rangle\ + 0.5
\langle H\sub{SB} \rangle$) between the bath with only single
excitations allowed $\langle E_{1}\rangle$ and the bath with two
simultaneous excitations $\langle E_{2} \rangle $ as a function of
time. The difference is calculated for a few coupling strengths.
The simulation time is $t=900$ fs and the number of bath modes is
$N=10 \ldots 40$. (Right panel) The population $P\sub{sim}$ of 1 and 2
simultaneous bath excitations is compared to the bath with two
simultaneous excitations. The solid lines refer to the weak coupling
limit ($\gamma^{-1} = 1630$ fs) and the dashed lines refer to  medium
coupling ($\gamma^{-1} = 163$ fs).}
\label{fig:energy_diff}
\end{figure}
%%%%%%%%%%%%%%%%%%%%%%%%%%%%%%%%%%%%%%%%%%%%%%%%%%%%%%%%%%%%%%%%%%%%%%%%%

First a restricted {\su} is applied, which limits the possible
system-bath correlations. The most extreme restriction includes
only single excitations. The results for the weak coupling case
($\gamma^{-1}=1630$ fs) are shown in Fig.~\ref{fig:weak}. For a
short period of time the energy relaxes with the same rate in the
two types of bath. However, after $t\sub{s}\approx500$ fs the rate
decreases and eventually the system energy becomes constant. It
should be pointed out, that the saturation time is not the
recurrence (Poincar\'{e}) time ($t\sub{s}<\tau\sub{rec}$). This is confirmed
by the fact that for time $t>t\sub{s}$ the overall energy transfer from
the bath back to the system is not complete. Calculating the
population of the bath modes shows that at $t>t\sub{s}$ most of the
system energy is transferred to very few (or even one) bath modes,
which are in resonance with the system's frequency. Modes which
are near to the resonance mode or modes become saturated and start to
transfer the excitation back to the system. A dynamic ``steady
state'' between the system and the bath is formed, where most of the modes
transfer energy back, while one (or very few) continue to absorb
energy from the system.

%%%%%%%%%%%%%%%%%%%%%%%%%%%%%72%%%%%%%%%%%%%%%%%%%%%%%%%%%%%%%%%%%%%%%%%%
\begin{figure}
\includegraphics[scale=0.5]{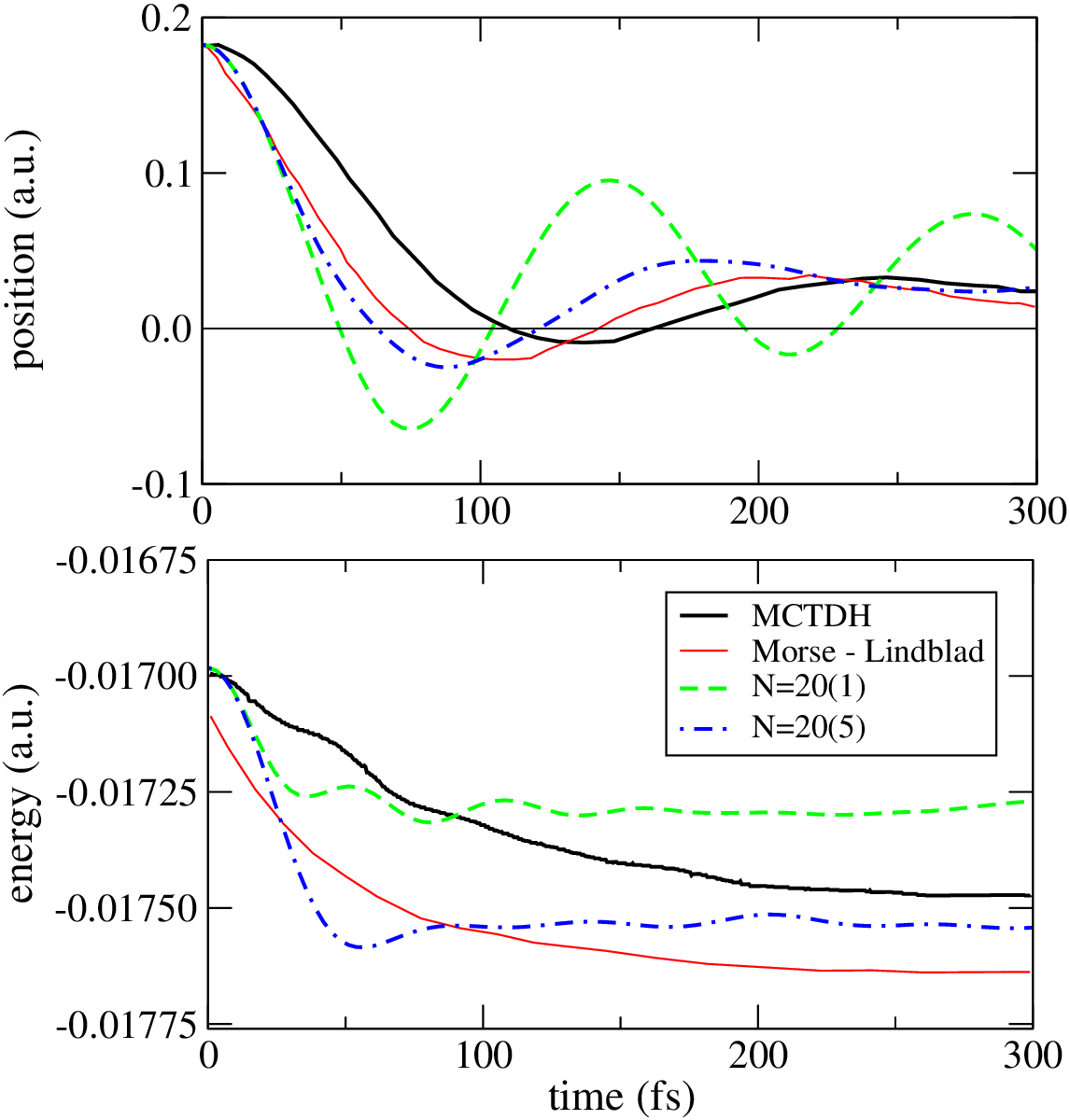}
\caption{\footnotesize  The energy relaxation (lower panel) and
damped oscillations of the average position (upper panel) of the
Morse oscillator in the strong coupling strength ($\gamma^{-1}=54$
fs). The bath parameters and the initial state are the same as in
the weak coupling calculations. The number of bath modes is
$N=20$. Solid lines refer to Nest and Meyer's MCTDH calculations
(adopted from Ref. \cite{meyer03}), where the thick lines are the
full-dimensional wave packet result and the thin lines refer to
the Morse-Lindblad model. Dashed lines refer to {\su} calculations
with single excitations. Dashed-dotted lines refer to a bath with
five excitations allowed simultaneously.} \label{fig:strong}
\end{figure}
%$$$$$$$$$$$$$$$$$$$$$$$$$$$$$$$$$$$$$$$$$$$$$$$$$$$$$$$$$$$$$$$$$$$$$$$$$$$$$

When the number of simultaneous excitations is increased to two,
the effect of saturation appears at a later stage ($t\sub{s}>2000$ fs).
The results become similar to those of Ref. \cite{meyer03} and the
values of the average position (see Fig.~\ref{fig:weak} (upper panel))
are nearly indistinguishable. We conclude that for the weak coupling
case, the bath that has two simultaneous excitations is
completely sufficient to reproduce the dynamics generated by {\it all }
simultaneous excitations for times up to 2 ps.

The relaxation dynamics for medium coupling are shown in
Fig.~\ref{fig:medium}. A saturation effect was obtained for the
bath restricted to single excitations. However, two simultaneous
excitations were sufficient to overcome this saturation and
converge the whole dynamics of the problem. A slight difference in
the energy relaxation rate of the two baths is identified. The TLS
bath causes stronger relaxation, but the results are still in good
agreement with those of Ref.~\cite{meyer03}. Since the initial
state is a function of $\Op{R}$ and the system-bath coupling
depends on $\Op{R}$ as well, the initial excitation influences the
effective strength of the coupling. If the initial displacement,
i.e. the initial excitation of the primary system, is decreased,
the saturation is postponed. We can then deduce that the
relaxation rate converges to the value of Ref.~\cite{meyer03}.
Combining the results of Figs.~\ref{fig:weak} and \ref{fig:medium}
leads to the conclusion that the differences between the two types
of bath in the weak and intermediate coupling regimes are caused
by the saturation of a few ``central'' modes in the spin bath.
This saturation is postponed if the bath includes more correlations.
For very weak coupling, these higher order system-bath correlations
become insignificant.

The problem of including all system-bath correlations is therefore
crucial in the medium and strong coupling regime.
Fig.~\ref{fig:energy_diff} shows the difference in the system
energy ($\langle \Op{H}\sub{S} \rangle\ + 0.5 \langle
\Op{H}\sub{SB} \rangle$) for two cases: a bath with only single
excitations and a bath in which two simultaneous excitations are
allowed. The calculations were made for different coupling
strengths. As the coupling strength is reduced, the difference
decreases. Thus in a very weak coupling limit, the TLS bath with
only single excitations (no system-bath correlations) becomes
sufficient to describe the dynamics for relatively long times. In
this limit the TLS bath  coincides completely with the harmonic
bath.

The issue of including system-bath correlations has also been addressed
in the MCTDH calculations. In Ref.~\cite{burghardt} the same system has
been studied with the G-MCTDH method (the MCTDH with Gaussian expansion
functions). Differences between the single-configurational (the
Hartree limit) and the multi-configurational descriptions (with an
increasing number of single particle functions) have been obtained
for the energy relaxation process. In these calculations at least four
single particle functions per resonant bath modes and two spf for
secondary modes were required to achieve convergence in the
relatively weak coupling limit ($\gamma^{-1}=500$ fs).

In the strong coupling regime (Fig.~\ref{fig:strong}) there is
considerable deviation between the two models. In the {\su} model
the energy relaxes faster and the oscillator is damped after a
single period. The relaxation rate obtained for the TLS bath is
closer to the one obtained by the Morse-Lindblad model (adapted
from Ref.~\cite{meyer03}). As expected convergence requires many
simultaneous excitations. For example, a bath consisting of
$N=20$ modes required at least five simultaneous excitations for
converging the energy relaxation dynamics.
%%%%%%%%%%%%%%%%%%%%%%%%%%%%%72%%%%%%%%%%%%%%%%%%%%%%%%%%%%%%%%%%%%%%%%%%

\subsection{The interaction between the bath modes.}
\label{subsec:interr}

The saturation of the TLS bath modes can be eliminated by allowing
energy exchange between bath modes. This is done by adding to the
bath Hamiltonian $\Op{H}\sub{B}$ of Eq.(\ref{eq:Htot}) the term:
\begin{equation}
\Op{H}\sub{int} =
\sum_{ij} \kappa_{ij} (\Op{\sigma}_i^\dagger \Op{\sigma}_j
+ \Op{\sigma}_j^\dagger \Op{\sigma}_i)\; ,
\label{eq:Hint}
\end{equation}
where the parameter $\kappa_{ij}(=\kappa^{*}_{ji})$ is the
interaction strength between two bath modes. The interaction can
be restricted to the nearest neighbors in energy by the condition
$\kappa_{ij}=0$ for $|i-j|>1$. The detailed algorithm of
applying Eq.(\ref{eq:Hint}) in the bit representation has been
described in Ref. \cite{ckoch} in the context of pure dephasing.

The term $\Op{\sigma}_i^\dagger \Op{\sigma}_j +
\Op{\sigma}_j^\dagger \Op{\sigma}_i$ describes a two
quasi-particle interaction within the bath. A qualitative picture
is based on an almost elastic exchange of energy between the two
nearest neighbor bath modes which are almost degenerate.
The process  is described by a creation of an excitation in one
mode at the expense of another and vice versa.

The new bath Hamiltonian including the interactions can be
diagonalized leading to:
\begin{equation}
\Op{\tilde{H}}\sub{B} =
\sum_i \tilde{\omega}_i \Op{\tilde{\sigma}}_i^\dagger \Op{\tilde{\sigma}}_i \; ,
\end{equation}
where $\tilde{\omega}_i$ are the eigenvalues of
$\Op{D}^\dagger(\Op{H}\sub{B}+\Op{H}\sub{int})\Op{D}$. In the new
basis of $\{\Op{\tilde{\sigma}}_i\}$ the system-bath interaction
term in Eq. (\ref{eq:Hsb}) is also modified. However for
sufficiently small $\kappa$ the eigenvalues of the bath change
only slightly, but the saturation effect is postponed to a much
later time.

%$$$$$$$$$$$$$$$$$$$$$$$$$$$$$$$$$$$$$$$$$$$$$$$$$$$$$$$$$$$$$$$$$$$$$$$$$$$$$$$$
\begin{figure}
\includegraphics[scale=0.5]{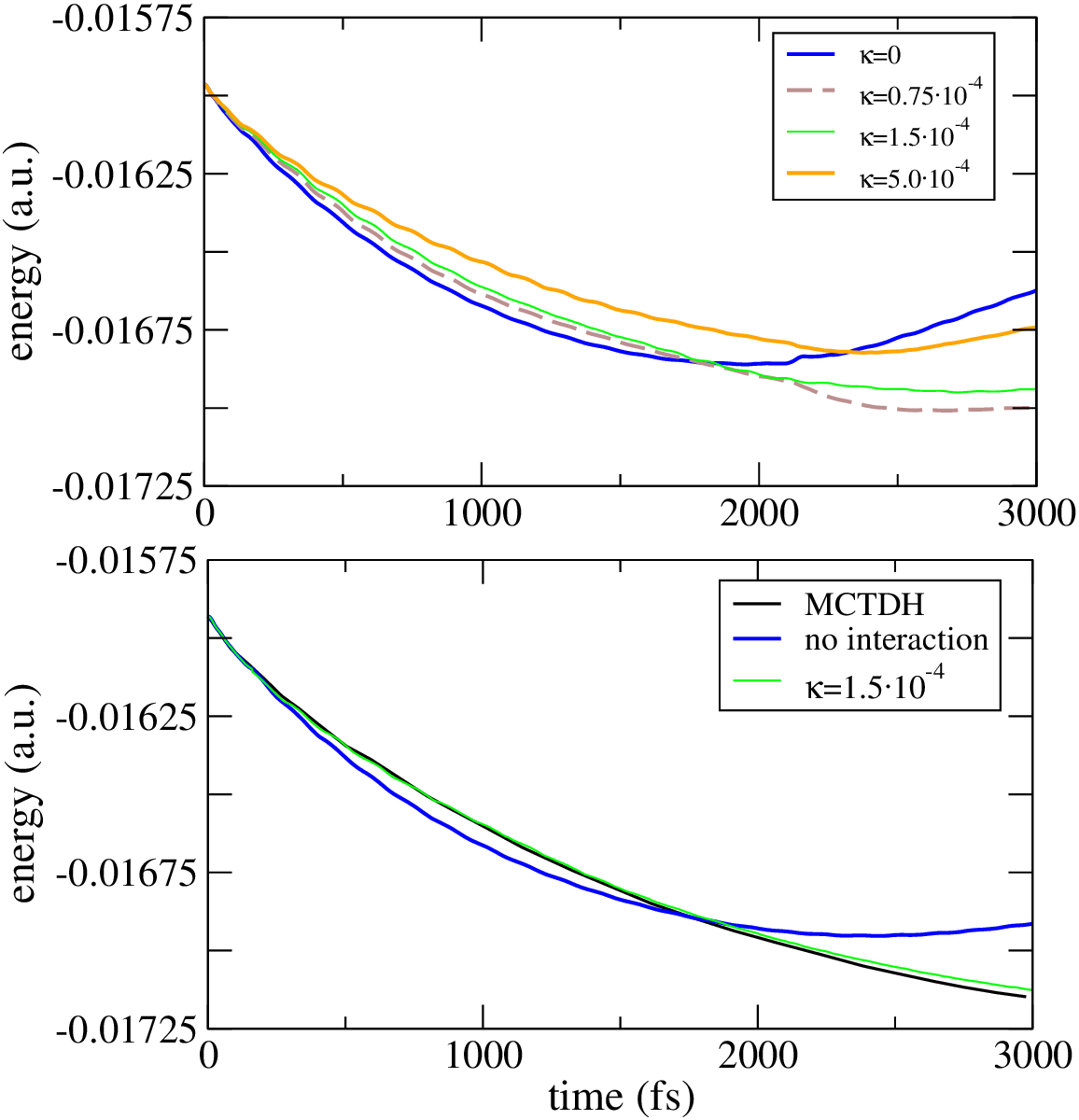}
\caption{\footnotesize The energy relaxation with interaction
between the bath modes is shown in the weak coupling limit
($\gamma^{-1} = 1630$ fs). The bath is assumed to be Ohmic with
cutoff frequency $\omega\sub{c}=2.9\cdot10^{-3}$ a.u. (Upper panel)
The influence of the parameter $\kappa$ is shown for a bath of
$N=40$ modes. (Lower panel) The energy relaxation is shown with
the optimal parameter of $\kappa=1.5\cdot10^{-4}$ (thin line).
The thick solid line refers to MCTDH calculations with a bath of
harmonic oscillators (\cite{meyer03}). The dashed line refers to {\su}
calculations with two simultaneous excitations allowed without
interaction between the bath modes. The bath consists of $N=60$
modes.}
\label{fig:inter}
\end{figure}
%$$$$$$$$$$$$$$$$$$$$$$$$$$$$$$$$$$$$$$$$$$$$$$$$$$$$$$$$$$$$$$$$$$$$$$$$$$$

Fig.~\ref{fig:inter} shows the influence of the interaction
between the bath modes on energy relaxation in the weak coupling
limit ($\gamma^{-1}=1630$ fs). The dynamics are calculated for a
relatively long period of 3 ps. For such a long time the saturation
effect is observed even for a bath with two simultaneous
excitations. Since the saturation time $t\sub{s}$ is determined mostly
by the saturation of the few modes close to resonance with the
subsystem, increasing the number of modes cannot prolong $t\sub{s}$.

Adding an interaction between the bath modes leads to slower decay
and delayed saturation (Cf Fig.~\ref{fig:inter}, upper panel).
This can be understood from the following considerations: the
interaction term, Eq.(\ref{eq:Hint}), describes the transport of
excitation from one bath mode to its nearest neighbor. Consequently,
$\kappa$ determines how quickly the excitation is transported away
from a TLS mode close to resonance with the primary system. On the
other hand, the interaction energy, i.e. the expectation value of
$\langle \Op{H}\sub{SB}\rangle$, depends on the population of the
primary system and of the bath modes close to it. If the population is
removed from those bath modes and ``diffuses'' all over the bath,
the interaction energy decreases and the decay becomes slower.
This explains the upper panel of Fig.~\ref{fig:inter} which shows
the energy relaxation for different values of $\kappa$.

An optimal value of $\kappa$ can minimize the differences between
the spin and the harmonic bath. To demonstrate the effect, the
calculations were carried out for $N=60$ bath modes and
$\kappa=1.5\cdot10^{-4}$ (Cf. the lower panel of Fig.
~\ref{fig:inter}). For this value of $\kappa$ the spectrum of the
bath was only slightly altered (less than 1$\%$). The energy
relaxation in this case is almost indistinguishable from the
results obtained in Ref. \cite{meyer03}.

\subsection{Correlated versus uncorrelated states.}
\label{subsec:correlation}
%$$$$$$$$$$$$$$$$$$$$$$$$$$$$$$$$$$$$$$$$$$$$$$$$$$$$$$$$$$$$$$$$$$$$$$$$$$
\begin{figure}
\includegraphics[scale=0.5]{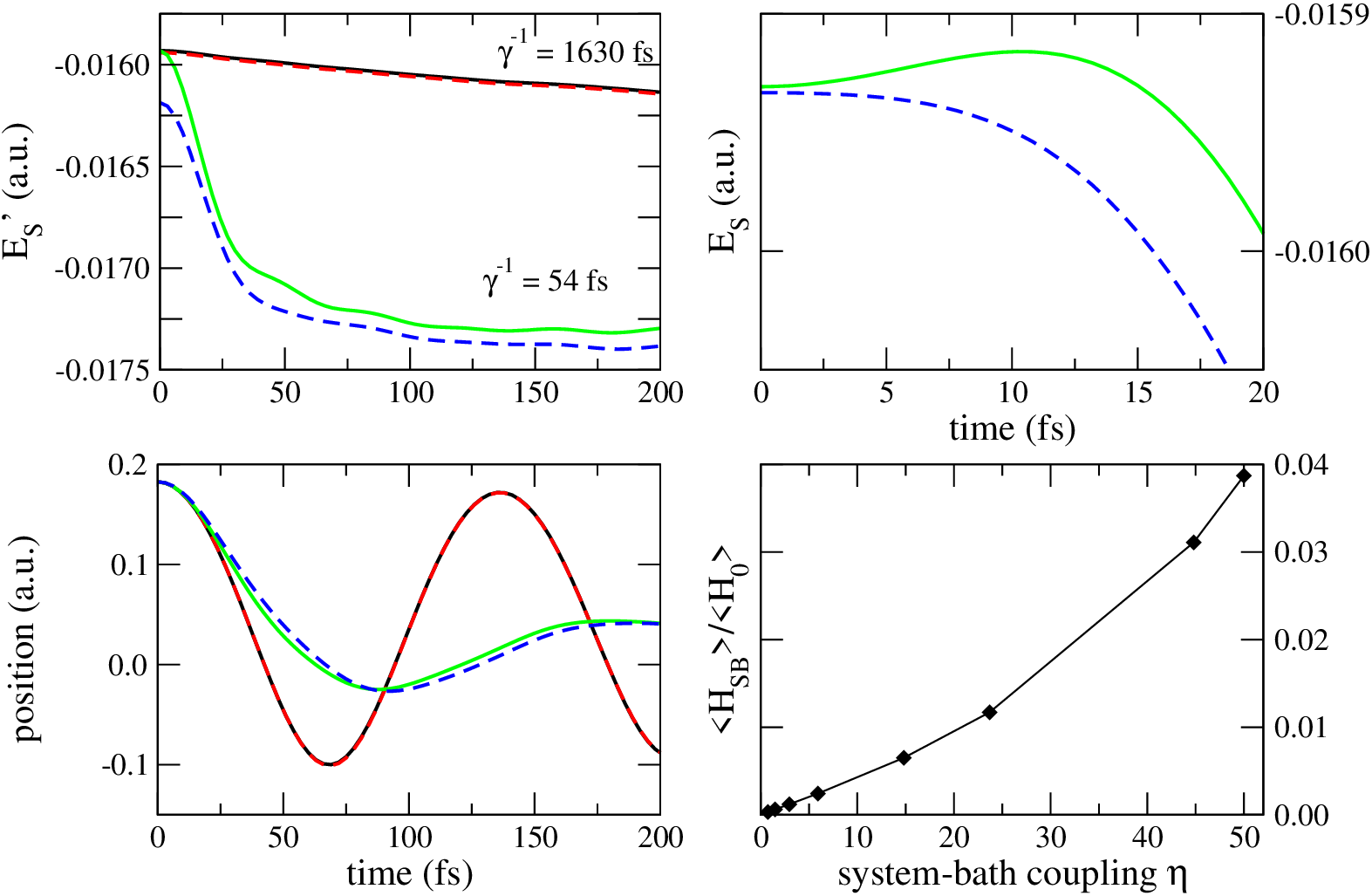}
\caption{\footnotesize Effect of initial correlations. The energy
relaxation and damped oscillations of the average position are
shown for the initially uncorrelated (solid lines) and correlated
(dashed lines) states. The dynamics for weak ($\gamma^{-1}=1630$
fs) and strong couplings ($\gamma^{-1}=54$ fs) are compared. A
bath consisting of $N=10$ modes with 2 and 5 simultaneous
excitations (for the weak and strong coupling, respectively) is
sufficient to obtain the converged results. (Left) The effective
subsystem energy (upper panel) ($E\sub{S}^{'}=\langle H\sub{S}
\rangle\ + 0.5 \langle H\sub{SB} \rangle$) and the expectation
value for the Morse coordinate (lower panel) are shown. (Right)
The bare subsystem energy (upper panel) $E\sub{S}=\langle H\sub{S}
\rangle$ in the strong coupling regime ($\gamma^{-1}=54$ fs) is
shown for the short time dynamics. The energy stored in the
system-bath coupling (lower panel) is calculated as a function of
the coupling constant $\eta=M\gamma$.} \label{fig:correl}
\end{figure}
%$$$$$$$$$$$$$$$$$$$$$$$$$$$$$$$$$$$$$$$$$$$$$$$$$$$$$$$$$$$$$$$$$$$$$$$$$$$$$

The widely used assumption of an initially uncorrelated
system-bath state is not consistent with most experimental
situations \cite{pechukas94,romero,geva}. The influence of 
initial correlations has been addressed in the context of 
the weak coupling approximation, where it appears as an additional
inhomogeneous term \cite{geva}.

A fully correlated initial state is easily obtained in
the {\su} method. Once the system-bath Hamiltonian $\Op{H}$ is
set, the correlated ground state can be determined by propagating
an initial guess wave function in imaginary time using $\Op{H}$
\cite{k42}.

The influence of initial correlations is shown in
Fig.~\ref{fig:correl}. The uncorrelated state is identical to that
of the previous calculations: the primary system is defined as a
shifted Gaussian wave packet, while the bath is not excited. For
the correlated initial state, the ground state of the total
system was calculated first. Then this ground state was displaced by the
shift operator in momentum space $\Op{D}= e^{-i R_0 \Op{k}}$ with
$R_0=2\tilde{R}$. The dynamics of the correlated state are
compared to that of the uncorrelated state for weak and strong
couplings ($\gamma^{-1}=1630$ fs and $\gamma^{-1}=54$ fs). The
dashed and solid lines in Fig.~\ref{fig:correl} correspond to the
initial state being correlated and uncorrelated, respectively.

The short-time dynamics differ for the correlated and uncorrelated
cases, since the correlations need to be built up in the
uncorrelated case \cite{suarez}. In the latter case an initial slippage
in the system energy can be observed before the reduced dynamics
appear to be  Markovian (right upper panel). This effect is insignificant
for weak coupling. Even for strong coupling, the differences
between the correlated and uncorrelated cases were found to be
very small. Apparently, the displacement is a stronger
"perturbation" than that caused by the correlations, i.e. the
displacement establishes a new initial state
\cite{MancalMayJCP01}.

\subsection{Decoherence.}
\label{subsec:dephasing}

Decoherence has become a popular term used to describe loss of
phase in coherent superpositions of quantum states due to
interaction with a bath. It is therefore natural to compare the
decoherence properties of the spin bath to those of the harmonic bath. The
first difficulty is that there are different approaches to the
definition of decoherence. Alicki \cite{alicki03} identifies
pure decoherence (dephasing) with the decay of the off-diagonal elements
of the density operator, which is not accompanied by dissipation.
He then argues that dephasing cannot be caused by a harmonic
oscillator bath with a coupling, which is linear in coordinates or
momenta.

Energy relaxation is also accompanied by loss of phase. For
comparison, we will consider decoherence as a process
caused by energy relaxation, which is characterized by a time
$T_1=\gamma^{-1}$. The decoherence effect will be illustrated in
terms of the dissipative dynamics of cat states, defined as a
superposition of two coherent states. The interaction with the
environment leads to decay of the coherences of such a
superposition on an extremely short time scale, usually much
shorter than the corresponding relaxation time scale
\cite{joos_book}. This process has been modeled using G-MCTDH by
\cite{burghardt}, and as a result can be used for a comparative
study.

The Wigner function of the cat state
consists of two Gaussians centered at $(\pm R_0,p_0)$ and an
interference term, which is centered at the origin. The
off-diagonal part of the density matrix in the coherent-state
basis, which contains information about quantum interferences
between the two components of the cat state, decays with the rate
$\gamma\sub{coh}$. In the Markovian limit the decay rate is
proportional to the square of the distance between the coherent
states. For zero temperature it is given by
\cite{haake85,rajagopal}:
\begin{equation}
\gamma\sub{coh} = \frac{\gamma M \omega_0 \delta^{2}}{2\hbar} \; .
\label{eq:drate}
\end{equation}
$\omega_0$ and $\delta$ are  parameters of the primary system
($\omega_0$ represents the frequency of the harmonic oscillator, and
$\delta$ is the separation distance between the coherent states).

The decoherence rate for a primary system coupled to a TLS bath is
calculated and compared to the calculations for a bath of harmonic
oscillators with the G-MCTDH method \cite{burghardt}. The
calculations are performed for a cat state in a harmonic
oscillator potential with $\omega_0=10^{-3}$ a.u. and $M=10^5$
a.u. The bath has the same parameters as for the previous
calculations with damping rates of $\gamma^{-1}=1630$ fs and
$\gamma^{-1}=500$ fs.

%%%%%%%%%%%%%%%%%%%%%%%%%%%%%%%%%%%%%%%%%%%%%%%%%%%%%%%%%%%%%%%%%%%
\begin{figure}
\includegraphics[scale=0.35]{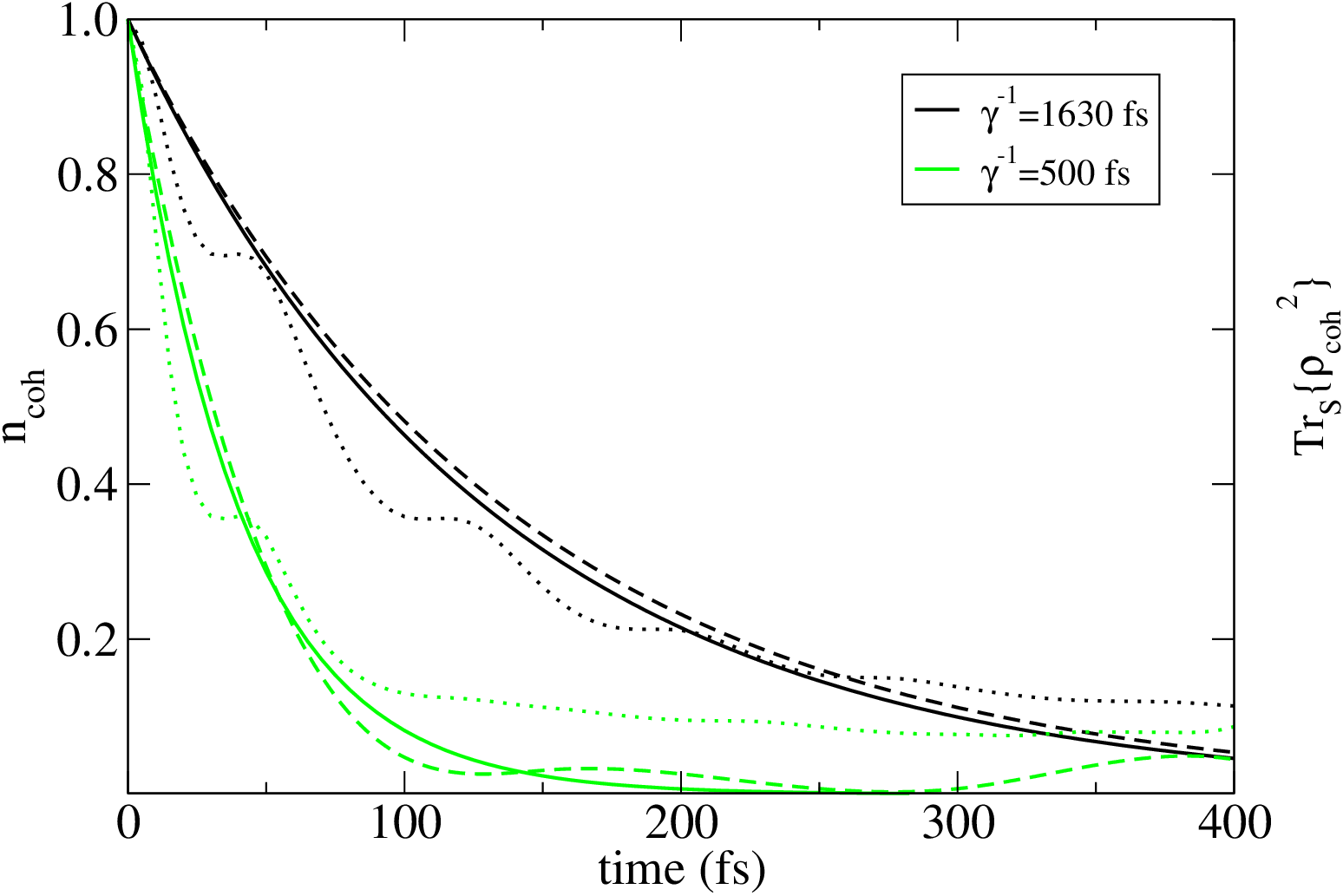}
\caption{\footnotesize The decoherence effect in terms of decay of
the coherence norm and off-diagonal elements in the energy
representation. The coherence norm $n\sub{coh}$ is shown as a
function of time for two different couplings, $\gamma^{-1} = 1630$
fs and $\gamma^{-1} = 500$ fs. In the golden rule limit,
off-diagonal elements of the reduced system density matrix
$\rho(R,R^{'})$ decay exponentially with the decoherence rates
$\gamma\sub{coh}^{-1} = 130$ fs and $\gamma\sub{coh}^{-1} = 40$ fs
(for the two given couplings, respectively). This decay is shown by the
full lines. The dashed lines refer to the calculated decay of the
decoherence norm. The dotted lines show the decay of $\Fkt{tr}_S
\{\Op{\rho}\sub{coh}^2\}$ in the energy representation.}
\label{fig:decoh}
\end{figure}
%%%%%%%%%%%%%%%%%%%%%%%%%%%%%%%%%%%%%%%%%%%%%%%%%%%%%%%%%%%%%%%%%

A quantitative measure of decoherence of the primary system is the
coherence norm used by Strunz et al. \cite{haake1}:
\begin{equation}
n\sub{coh}(t) = \Fkt{tr}_S\left\lbrace \Op{\rho}\sub{S}\up{coh}(t)
\Op{\rho}\sub{S}\up{coh\dagger}(t)\right\rbrace \; ,
\label{eq:coh_norm}
\end{equation}
where $\Op{\rho}\sub{S}\up{coh}$ refers to off-diagonal elements of
the subsystem reduced density matrix in the basis of coherent
states.

Since decoherence is a basis-dependent phenomenon, one can ask if
it can also be measured in the basis of the eigenstates of the
system Hamiltonian $\Op{H}\sub{S}$.  The question arises, whether
these eigenstates form a pointer basis \cite{zurekpaz} - the basis
with respect to which off-diagonal elements in the reduced density
operator disappear due to decoherence.
To perform this test the system density operator
$\Op{\rho}\sub{S}(t)= \Fkt{tr}_B \{{\Op \rho}\}$,
which has been calculated in the coordinate basis is
transformed to the basis of the $\Op{H}\sub{S}$ eigenstates.
Decomposing such a state to a dynamical and a static part leads
to \cite{banin94}:
\begin{equation}
\Op {\rho}\sub{S}(t)=\Op{\rho}\sub{coh}(t)  + C^2\Op {\rho}\sub{S}\up{eq}
\label{eq:rho_e}
\end{equation}
where $\Op {\rho}\sub{S}\up{eq}$ is the equilibrium stationary
system density operator and $C^2$ is an overlap functional given
by
\begin{equation}
C^2 = \Fkt{tr}_S \{\Op{\rho}\sub{S}(t) \cdot
\Op{\rho}\sub{S}\up{eq}\}/\Fkt{tr}_S\{{\Op{\rho}\sub{S}\up{eq}}^2\}
~~.\label{eq:csquare}
\end{equation}
$\Op{\rho}\sub{coh}(t)$ in Eq.(\ref{eq:rho_e}), has no diagonal elements
in the energy representation and is therefore traceless. Thus the
decoherence effect is measured by the decay of $\Fkt{tr}_S
\{\Op{\rho}\sub{coh}^2\}$.
\\
Fig.~\ref{fig:decoh} shows the decay of the coherence norm
$n\sub{coh}$ and $\Fkt{tr}_S\{\Op{\rho}\sub{coh}^2\}$ for two
different coupling strengths (both are weak). The thick lines
refer to a simple exponential decay predicted by Eq.(\ref{eq:drate})
for a harmonic bath  and confirmed by Ref. \cite{burghardt}. The
dashed lines refer to the calculated decay of $n\sub{coh}$, which
is in good agreement with the prediction. The decay of $\Fkt{tr}_S
\{\Op{\rho}\sub{coh}^2\}$ (the dotted lines) has almost the same
rate at a relatively short time. However, the off-diagonal elements
of $\Op{\rho}\sub{S}$ in the energy representation do not decay
strictly to zero. Therefore, in this case, the system energy
eigenstates cannot be considered as a pointer basis
\cite{zurekpaz}.

\subsection{Entanglement.}
\label{subsec:entanglement}

%%%%%%%%%%%%%%%%%%%%%%%%%%%%%%%%%%%%%%%%%%%%%%%%%%%%%%%%%%%%%%%%%%%%%%%%%
\begin{figure}
\includegraphics[scale=0.5]{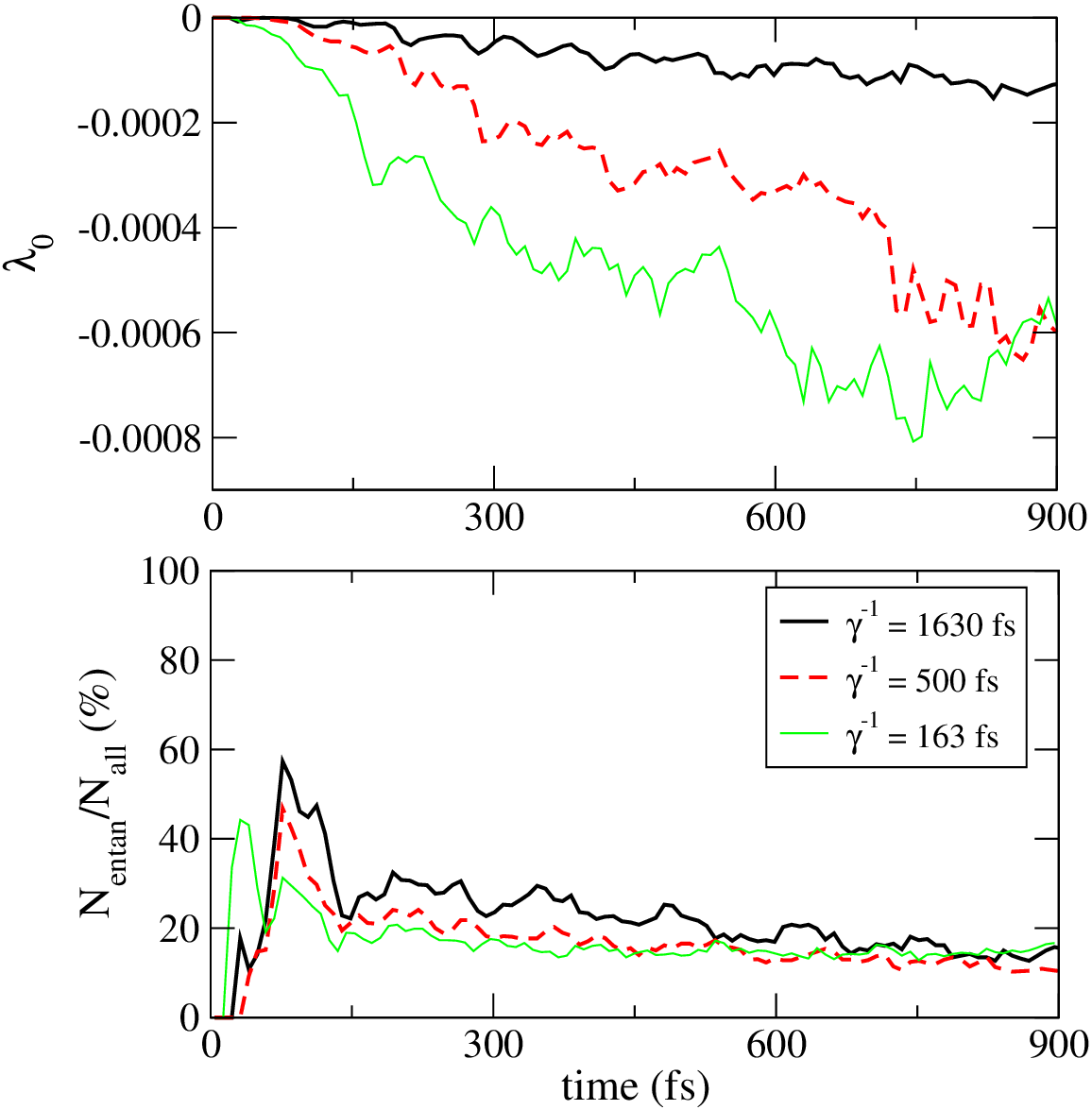}
\caption{\footnotesize Measurement of entanglement between the
bath modes as a function of time. (Upper panel) The smallest
eigenvalue of the partial transposition $\rho\up{T\sub{j}}$ of the
reduced density matrix for pair of the bath modes ($i,j$) is
calculated according to Appendix \ref{sec:horod_parameter}. The
negative eigenvalues are averaged over all possible combinations
of the bath modes. (Lower panel) The relative number of entangled
pairs of the bath modes as a function of time. The calculations
are performed for three different system-bath coupling strengths
($\gamma^{-1}=1630, 500, 163$ fs). The bath consisting of $N=40$
modes with two simultaneous excitations allowed is used in all
calculations.}
\label{fig:entanglement}
\end{figure}
%%%%%%%%%%%%%%%%%%%%%%%%%%%%%72%%%%%%%%%%%%%%%%%%%%%%%%%%%%%%%%%%%%%%%%%%

%%%%%%%%%%%%%%%%%%%%%%%%%%%%%%%%%%%%%%%%%%%%%%%%%%%%%%%%%%%%%%%%%%%%%%%%%
\begin{figure}
\includegraphics[scale=0.35]{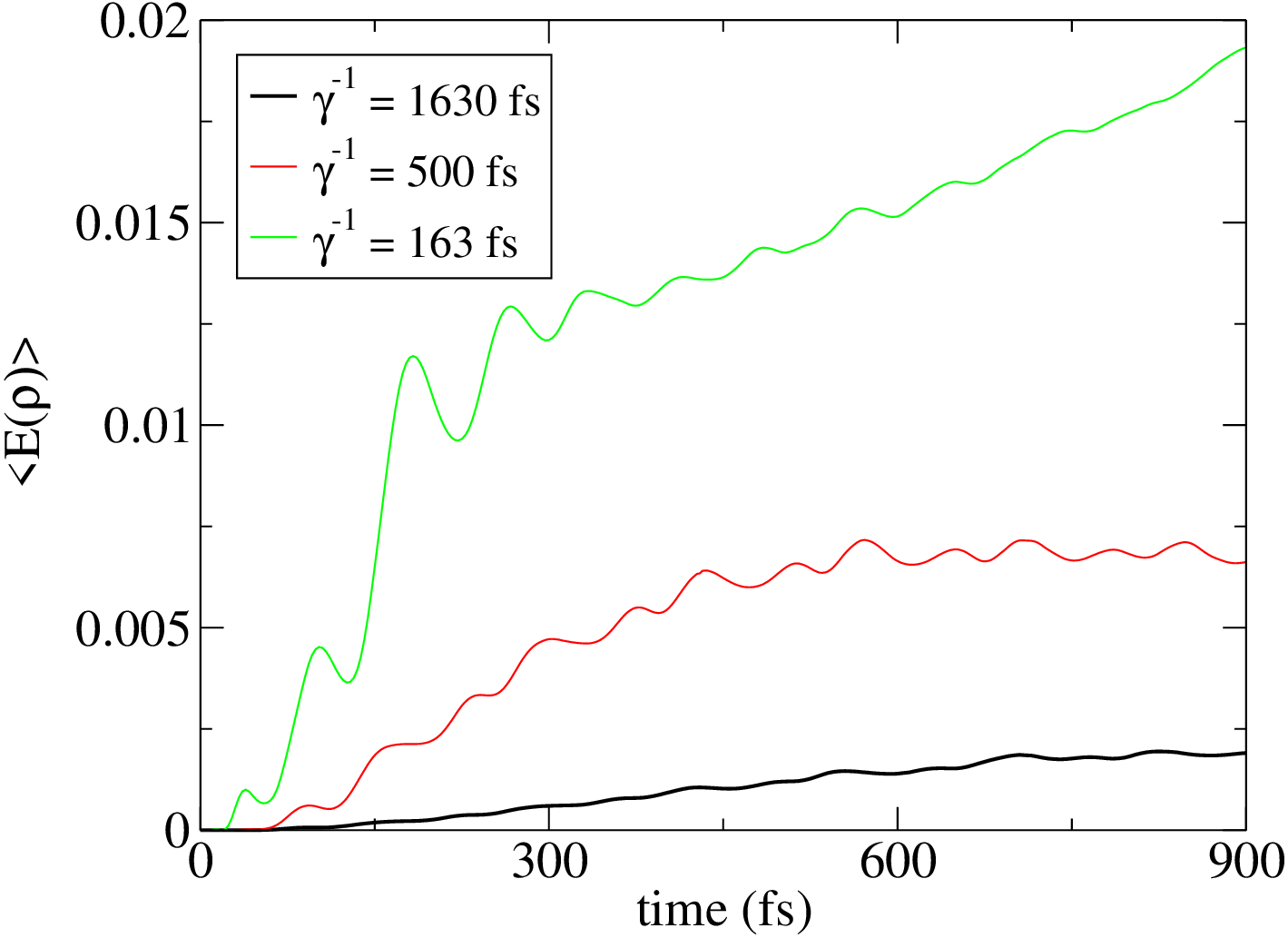}
\caption{\footnotesize The entanglement of formation $E(\rho)$ is
calculated for a couple of the bath modes ($i,j$). The average
over all possible pairs is shown as a function of time. The
calculations are performed for three different system-bath
coupling strengths ($\gamma^{-1}=1630, 500, 163$ fs). The bath
consisting of $N=40$ modes with two simultaneous excitations
allowed is used in all calculations. }
\label{fig:concurrence}
\end{figure}
%%%%%%%%%%%%%%%%%%%%%%%%%%%%%72%%%%%%%%%%%%%%%%%%%%%%%%%%%%%%%%%%%%%%%%%%

Entanglement between two quantum states is a manifestation of
additional quantum correlation. For example entanglement between
the system and the bath means $\Op \rho \neq
\Op{\rho}\sub{S}\otimes\Op{\rho}\sub{B}$. In a dissipative
environment it is expected that initial entanglement between parts
of the system are lost leading to decoherence
\cite{zurek,joos}. In addition a bath can also provide an indirect
interaction between totally decoupled parts of the primary system
and entangle them \cite{braun,benatti}.

The difference between the harmonic and the spin baths should be
manifested in another type of entanglement - quantum correlations
between different bath modes. A system interacting with  the spin
bath, can induce entanglement between two spin modes, which are not 
directly interacting with each other. In the harmonic bath on the
other hand, a system linearly coupled to different modes is not able 
to entangle those modes (see Appendix \ref{sec:equations}).

Peres \cite{peres} and Horodecki et al. \cite{horodecki} have
provided a criterion, based on partial transposition, to determine
whether a given mixed state of two subsystems is entangled (cf
Appendix \ref{sec:horod_parameter}). Since the criterion is
defined only for two coupled TLS, the study of entanglement is
limited to {\it two} bath modes (i) and (j). The density operator
of any two bath modes $\Op{\rho}\sub{ij}$ is obtained as a partial
trace of $\Fkt{tr}\sub{k\neq i,j}\{\Op{\rho}\sub{B}\}$ over the
rest of the bath modes, where the density operator of the bath is
$\Op{\rho}\sub{B}=\Fkt{tr}_S\{\Op{\rho}\}$. The procedure checks
whether the partial transposition of $\Op{\rho}\sub{ij}$ with
respect to one of the modes has negative eigenvalues. The smallest
eigenvalue $\lambda_0$ of the partial transpose matrix
$\rho\up{T\sub{j}}$ constitutes the criteria. Then the eigenvalues
with $\lambda_0<0$ are averaged over all {\it entangled} pairs of
bath modes.

In Fig.~\ref{fig:entanglement} the averaged parameter $\lambda_0$
is shown as a function of time for three different coupling
strengths. The entanglement calculations were based on converged
results obtained for a bath of $N=40$ modes. This was sufficient
to a time scale of 900 fs, for all three system bath coupling
strength considered. Since at $t=0$ the bath is not excited, there are no
entangled bath modes, therefore $\lambda_0=0$ for all pairs. As $t$
increases $\lambda_0$ becomes negative for some of the pairs of
the bath modes, meaning that these modes become entangled. As time
progresses, the number of entangled pairs  saturates for all
three couplings (Cf Fig.~\ref{fig:entanglement}, lower panel).
Therefore, the increase in the absolute value of $\lambda_0$ is
related primarily to the growth in population of the entangled
modes. The maximum in the number of entangled modes for an early
time may be associated with the creation of higher-order
entanglement terms where three or more simultaneous excitations
become important. Such higher-order entanglement is not captured
by the Peres-Horodecki parameter.

To characterize the degree of entanglement we use an additional 
measure, the entanglement of formation introduced by Wootters et al.
\cite{bennett,hill,wootters} (Cf. Appendix
\ref{sec:horod_parameter}). For any $2\otimes2$ mixed state this
quantity varies from zero (separable states) to one 
(maximally entangled states). Our results (Cf. Fig.~\ref{fig:concurrence})
are obtained by averaging the entanglement of formation $E(\rho\sub{ij})$ 
(for two bath modes $i$ and $j$) over all possible pairs of modes. 
The dynamics of $\langle E(\rho)\rangle$ is similar to those of the 
partial transpose parameter. It should be noted that the growth of
entanglement shown in
Figs.~\ref{fig:entanglement}-\ref{fig:concurrence} is exclusive to
the spin bath.

\section{Summary and Conclusions}
\label{sec:conclusions}

The similarities and differences of the relaxation dynamics of a
primary system coupled to a spin or to a harmonic bath have been
analyzed. The study was facilitated by the ability to obtain
converged numerical results for finite period in time. In both cases
this task becomes possible by employing a large but finite number of
bath modes and controlling the degree of correlation.

\subsection{The similarities}

For all cases studied the extremely short time dynamics was
identical. This period represents the inertial response of the
bath and is characterized by a zero derivative of the energy
at the initial time $t=0$. This non-Markovian dynamical
evolution, "the slippage", is quite short and in many cases it can be
ignored. The initial dynamics is closely related to the issue of
the choice of the initial state. Preferably it should represent
equilibrium system bath correlation and not be a product state.
The {\su} method allows to create such a fully correlated initial
state of the system-bath entity. However, in the present model
differences in the dynamics between correlated and uncorrelated
states seem to be insignificant, even in the strong coupling case.
It is expected that the same phenomena would be observed in the
harmonic bath.

For weak system-bath coupling the dynamics induced by both
baths are also similar. This is a numerical confirmation that in
the weak coupling limit the harmonic bath can be mapped to the
spin bath \cite{caldeira93,makri98,tsonchev}. In addition in this
limit the spin bath converges with only single excitations of the
bath modes meaning that the system and bath are almost
disentangled. This fact is consistent with the convergence to the
Markovian limit \cite{lindblad98}.

The decoherence properties of the harmonic and spin baths as
determined by the loss of phase of cat states are found to be
quite similar. This result is somewhat surprising since the
ergodic properties of the two baths are different. To rationalize,
one should notice that the coherence in cat states composed of a
superposition of two coherent states in a single mode does not
represent entanglement. Therefore, this phase loss does not 
characterize decoherence in accordance with Alicki's 
notion \cite{alicki03}. Moreover when a pure dephasing term was 
added it was found that it did not erode the phase coherence between 
cat states \cite{jiri}. We conclude that the decoherence properties 
of the two baths still require a further study.

\subsection{The differences}

The spin and harmonic baths begin to deviate when the initial
excitation of the primary system is increased. This difference is
observed for excitations where the dynamics generated by the
harmonic bath is still Markovian. The first indication of
differences is the necessity of two simultaneous bath excitation 
to converge the spin bath. For larger time periods, the spin bath
saturates limiting the ability to assimilate the system's energy.
The conclusion is that the limit of weak coupling is
more restrictive in the spin bath case. The effect of saturation
can be reduced if the bath Hamiltonian includes a mode-mode
coupling term. This term causes diffusion of excitation between
the modes, spreading the excitation over a greater number of bath
modes. Thus bath modes, which are relatively far from resonance
with the primary system become populated and the saturation is
suppressed. Practically, this allows to increase the convergence 
timescale of the spin bath.

In the medium coupling regime there is an overall good agreement
between the two models. The spin bath, however, causes stronger
relaxation, a fact, which becomes even more visible in the case of
strong coupling. In this regime the deviations between the two
baths become significant.

The possibility of entanglement of bath modes mediated by the
primary system is a major conceptual difference between the spin
and harmonic baths. In the spin bath after a short initial period
where only single excitations are excited, entanglement between
pairs of spin sets in with what seems as an exponential growth. At
later times the pair entanglement is replaced by higher order
terms and the pair entanglement saturates. All these correlations
are absent from the harmonic bath nevertheless the dynamics of the
the primary systems are not very different except for extremely
strong coupling. It could be possible that the in the present
model the primary system and the system bath coupling terms are
oversimplified. In addition the simulations should be extended to
finite temperatures, where different behavior of harmonic and spin
baths is expected except for the weak coupling regime
\cite{tsonchev,caldeira93,golosov}. The recently proposed random
phase method \cite{gelman} may be used to extend the above models
to finite temperature applications.

The present study is an important step in establishing the {\su}
method as a practical simulation tool. The elucidation of the
system-bath dynamics allows to tailor a simulation package in
particular for ultrafast dynamical processes.

\begin{acknowledgments}
We thank Roi Baer for his help and useful discussions.
This research was supported by  the  German-Israel Foundation
(GIF). The Fritz Haber Center is supported by the Minerva
Gesellschaft f\"{u}r die Forschung, GmbH M\"{u}nchen, Germany.
\end{acknowledgments}

\appendix
\section{Harmonic bath vs spin bath.}
\label{sec:equations}

The differences between harmonic and spin baths can be
illuminated by studying the simple system of a spin-$\frac{1}{2}$
coupled to a bath. For the harmonic bath the total Hamiltonian in
second quantization is given by:
\begin{equation}
\Op{H} = \frac{1}{2}\Omega \hat{\sigma}_{z} + \sum_{j}\omega_{j}
\Op{b}_{j}^{\dag} \Op{b}_{j} + \sum_{j} \lambda_{j} (\Op{b}_{j} +
\Op{b}_{j}^{\dag})(\hat{\sigma}_{+} + \hat{\sigma}_{-}) \; .
\label{eq:ham}
\end{equation}
Thus the Heisenberg equations of motion for the system operators
are:
\begin{equation}
\frac{d}{dt}\hat{\sigma}_{\pm}
=\frac{1}{i}[\hat{\sigma}_{\pm},\Op{H}]= \pm i\Omega
\hat{\sigma}_{\pm} - i \hat{\sigma}_{z}\sum_{j} \lambda_{j}
(\Op{b}_{j} + \Op{b}_{j}^{\dag}) \; .
\label{eq:motion}
\end{equation}
Similarly, the equations of motion for the bath operators are:
\begin{equation}
\frac{d}{dt}\Op{b}_{j} = -i\omega_{j} \Op{b}_{j} - i\lambda_{j}
(\hat{\sigma}_{+} + \hat{\sigma}_{-}) \; .
\end{equation}
Since the annihilation and creation operators $\Op{b}_{j}$ and
$\Op{b}_{j}^{\dag}$,  satisfy the standard Bose commutation
relation, $[\Op{b}_{j},\Op{b}_{j'}^{\dag}]=\delta_{j,j'}$, a closed
set of equations is obtained for each of the independent bath
modes.
\\
Now, let us consider a different model, where the primary system is
coupled to a bath of spins-$\frac{1}{2}$. The total Hamiltonian may
be written in the form:
\begin{equation}
\Op{H} = \Omega \hat{\sigma}_{z}^{0} + \frac{1}{2}
\sum_{j}\omega_{j} \Op{\sigma}_{z}^{j} - \frac{1}{2} \sum_{j}
\lambda_{j} \hat{\sigma}_{+}^{0}\Op{\sigma}_{-}^{j} + h.c. \; ,
\end{equation}
where $\Op{\sigma}_{x,y,z}$ designates the set of Pauli operators and
$\Op{\sigma}_{\pm}^{i}= \Op{\sigma}_{x}^{i} \pm
\Op{\sigma}_{y}^{i}$ are the usual ladder operators. For simplicity, 
we consider a system (0) consisting of a spin-$\frac{1}{2}$
which interacts with a pair (1,2) of spins. Thus, the Hamiltonian
of the whole system reads:
\begin{equation}
\Op{H} = \Omega \hat{\sigma}_{z}^{0} + \frac{1}{2}(\omega_{1}
\Op{\sigma}_{z}^{1} + \omega_{2} \Op{\sigma}_{z}^{2})-
\frac{\lambda_{1}}{2} (\hat{\sigma}_{+}^{0}\Op{\sigma}_{-}^{1} +
\hat{\sigma}_{-}^{0} \Op{\sigma}_{+}^{1}) - \frac{\lambda_{2}}{2}
(\hat{\sigma}_{+}^{0}\Op{\sigma}_{-}^{2} + \hat{\sigma}_{-}^{0}
\Op{\sigma}_{+}^{2}) \; .
\label{eq:spin_ham}
\end{equation}
The Heisenberg equations of motion for the system operator are:
\begin{equation}
\frac{d}{dt}\hat{\sigma}_{\pm}^{0} = \pm i (2 \Omega
\hat{\sigma}_{\pm}^{0} + \frac{\lambda_{1}}{2}
\hat{\sigma}_{z}^{0} \Op{\sigma}_{\pm}^{1} + \frac{\lambda_{2}}{2}
\hat{\sigma}_{z}^{0} \Op{\sigma}_{\pm}^{2}) \; ,
\end{equation}
and the equation of motion for the bath operator reads:
\begin{equation}
\frac{d}{dt} \Op{\sigma}_{\pm}^{1,2} = \pm i (\omega_{1,2} \Op{\sigma}_{\pm}^{1,2}
 + \frac{\lambda_{1,2}}{2} \sigma_{\pm}^{0} \Op{\sigma}_{z}^{1,2})\; .
\end{equation}
The commutation relations for spin operators are different from
those of bosons:
\begin{equation}
[\Op{\sigma}_{i},\Op{\sigma}_{j}]= - 2i \Op{\sigma}_{k} \; ,
\end{equation}
which makes the set of the equations above non-closed. After some
algebra, the equation for the bath operators becomes:
\begin{equation}
\frac {d}{dt} (\hat{\sigma}_{z}^{0} \Op{\sigma}_{\pm}^{1,2}) = \pm
i \omega_{1,2} \hat{\sigma}_{z}^{0} \Op{\sigma}_{\pm}^{1,2} \pm i
\frac{\lambda_{1}}{2} \hat{\sigma}_{\pm}^{0} + i
\lambda_{2}(\hat{\sigma}_{+}^{0} \Op{\sigma}_{\pm}^{1,2}
\Op{\sigma}_{-}^{2,1} - \hat{\sigma}_{-}^{0}
\Op{\sigma}_{\pm}^{1,2} \Op{\sigma}_{+}^{2,1}) \; .
\end{equation}
The triple correlations of the type $\hat{\sigma}_{+}^{0}
\Op{\sigma}_{\pm}^{1} \Op{\sigma}_{-}^{2}$ is a manifestation of
the build-up of {\it quantum entanglement} - a specific correlation
between different modes, which has no analogy in classical
physics. These correlations make a difference between the spin
bath and the harmonic oscillator one, since the latter does not
have quantum correlations between the bath modes.
\label{comparison}

\section{Entanglement between different bath modes.}
\label{sec:horod_parameter} In order to check whether the reduced
{\it two-system} density matrix $\Op{\rho}$ is entangled,
first we will use the partial transposition criterion
proposed by Peres \cite{peres} and Horodecki et al.
\cite{horodecki}. A mixed state described by density matrix
$\Op{\rho}$ is non-separable (and therefore cannot be written as a
product state of two subsystems (i) and (j),
$\Op{\rho}=\Op{\rho}\sub{i}\otimes \Op{\rho}\sub{j}$), iff the
partial transposition of $\Op{\rho}$ with respect to one of the
two subsystems has negative eigenvalues. The partial transpose
$\Op{\rho}\up{T\sub{j}}$ is obtained by transposing in a matrix
representation of $\Op{\rho}$ only those indices corresponding to
subsystem (j), i.e.
$\rho\up{T\sub{j}}\sub{m\mu,n\nu}=\rho\sub{m\nu,n\mu}$. The
following notation for matrix elements of a composite system is
used:
\begin{equation}
\rho\sub{m\mu,n\nu}=\left \langle e_m \otimes f_{\mu} |\Op{\rho}| e_n
\otimes f_{\nu} \right \rangle \; ,
\end{equation}
where ${e_m}$ and ${f_{\mu}}$ denote the arbitrary orthonormal
bases in Hilbert space describing the first (i) and second (j)
system, respectively.
\\
Checking the positivity of the partial transpose is equivalent to
checking the signs of the eigenvalues of $\Op{\rho}\up{T\sub{j}}$ 
or ,alternatively, the signs of the following determinants:
\begin{equation}
W_1=\rho\up{T_j}\sub{11,11}\rho\up{T_j}\sub{22,22}
-\rho\up{T_j}\sub{11,22}\rho\up{T_j}\sub{22,11}, \; \;
W_2=\rho\up{T_j}\sub{12,12}\rho\up{T_j}\sub{21,21}
-\rho\up{T_j}\sub{12,21}\rho\up{T_j}\sub{21,12}. \label{eq:horod}
\end{equation}
In the case when one of the above determinants is negative, the
state $\Op{\rho}$ is non-separable, and hence there is
entanglement between the two subsystems $\Op{\rho}\sub{i}$,
$\Op{\rho}\sub{j}$.

Another entanglement measure for a mixed state of two
spin-$\frac{1}{2}$ particles is the entanglement of formation
\cite{bennett,hill,wootters}. Explicitly, for the reduced
two-system density matrix $\Op{\rho}$ the entanglement of
formation is defined by
\begin{equation}
E(\rho)=h\left( \frac{1}{2}\left[1+\sqrt{1-\mathcal{C}(
\rho)^2}\right]\right) \;,
\end{equation}
where $h$ is the binary entropy function
$h(x)=-x\log_{2}{x}-(1-x)\log_{2}(1-x)$ and $\mathcal{C}(\rho)$ is
the concurrence. The concurrence is calculated in the following way:
first we define the "spin-flipped" density matrix to be
\begin{equation}
\widetilde{\rho}=(\sigma_y\otimes\sigma_y)\rho^{*}
(\sigma_y\otimes\sigma_y) \; ,
\end{equation}
where the asterisk denotes complex conjugation of $\Op{\rho}$ in
the standard basis
$\left \{ |00\rangle,|01\rangle,|10\rangle,|11\rangle,\right \}$ and
$\sigma_y$ expressed in the same basis is the matrix
\begin{equation}
\sigma_y=\left( {\begin{array}{cc} 0 & -i \\
i & 0 \end{array}} \right) \; .
\end{equation}
As both $\rho$ and $\widetilde{\rho}$ are positive operators, it
follows that the product $\rho\widetilde{\rho}$, though non-Hermitian,
also has only real and non-negative eigenvalues. Let the square
roots of these eigenvalues, in decreasing order , be $\lambda_1$,
$\lambda_2$, $\lambda_3$, and $\lambda_4$. Then the concurrence of
the density matrix $\rho$ is defined as
$\mathcal{C}=\max\{\lambda_1-\lambda_2-\lambda_3-\lambda_4,0\}$. It
should be noted that $\mathcal{C}=0$ corresponds to an unentangled
state, while $\mathcal{C}=1$ - to a completely entangled state and
the entanglement of formation $E$ is a monotonically
increasing function of $\mathcal{C}$.

\clearpage
\bibliography{morse}

\end{document}